\documentclass[journal]{IEEEtran}

\ifCLASSINFOpdf
\else
\fi

\usepackage[cmex10]{amsmath}
\usepackage{amsmath, amssymb}

\usepackage{ dsfont }
\usepackage{url}
\usepackage{cancel}
\usepackage{ mathrsfs }
\usepackage{theorem}
\usepackage{graphicx}

\usepackage{soul,xcolor}

\usepackage{todonotes}

\DeclareFontEncoding{LGR}{}{}
\DeclareTextSymbol{\~}{LGR}{126}

\theoremstyle{remark}
\theoremheaderfont{\normalfont\bfseries}
\theorembodyfont{\normalfont}

\usepackage[short]{optidef}

\usepackage{stackengine}
\usepackage{scalerel}
\usepackage{enumitem}
\usepackage{ gensymb }
\usepackage{subcaption}
\usepackage[numbers,sort&compress]{natbib}
\usepackage{mathtools}
\usepackage{booktabs}

\usepackage[subtle]{savetrees}
\allowdisplaybreaks[4]
\renewcommand{\baselinestretch}{1}
\usepackage[font=small,skip=2pt]{caption}

\renewcommand{\IEEEbibitemsep}{0pt plus 2pt}
\makeatletter
\IEEEtriggercmd{\reset@font\normalfont\fontsize{3.9pt}{3.40pt}\selectfont}
\makeatother
\IEEEtriggeratref{1}
\begin{document}
\bstctlcite{IEEEexample:BSTcontrol}
\setstcolor{red}
%
\title{A Model-Predictive Control Method for Coordinating Virtual Power Plants and Packetized Resources, with Hardware-in-the-Loop Validation}
%
%
%

 \author{Mahraz~Amini,~\IEEEmembership{}
       Adil~Khurram,~\IEEEmembership{}
         Andrew~Klem,~\IEEEmembership{}
         Mads~Almassalkhi,~\IEEEmembership{}
       and Paul D.~H.~Hines~\IEEEmembership{}
       
       \thanks{“This work was supported by the U.S. Department of Energy’s Advanced Research Projects Agency - Energy (ARPA-E) award DE-AR0000694. The authors are with the Department of Electrical and Biomedical Engineering at The University of Vermont, Burlington, VT 05405, USA.“}
}

\maketitle

\begin{abstract}
In this paper, we employ a bi-level control system to react to disturbances and balance power mismatch by coordinating distributed energy resources (DERs) under packetized energy management.
Packetized energy management (PEM) is a novel bottom-up asynchronous and randomizing coordination paradigm for DERs that guarantees quality of service, autonomy, and privacy to the end-user. A hardware-in-the-loop (HIL) simulation of a cyber-physical system consisting of PEM enabled DERs, flexible virtual power plants (VPPs) and transmission grid is developed in this work. 
A predictive, energy-constrained dispatch of aggregated PEM-enabled DERs is formulated, implemented, and validated on the HIL cyber-physical platform. The energy state of VPPs, composed of a fleet of diverse DERs distributed in the grid, depending upon the distinct real-time usage of these devices. The experimental results demonstrate that the existing control schemes, such as AGC, dispatch VPPs without regard to their energy state, which leads to unexpected capacity saturation. By accounting for the energy states of VPPs, model-predictive control (MPC) can optimally dispatch conventional generators and VPPs to overcome disturbances while avoiding undesired capacity saturation. The results show the improvement in dynamics by using MPC over conventional AGC and droop for a system with energy-constrained resources. \end{abstract}

%
\IEEEpeerreviewmaketitle



\section{Introduction}
%
%
%
%

The drive to reduce greenhouse gas emissions and declining capital costs are precipitating rapid increases in wind and solar generation capacity. Despite their low emissions profile, wind and solar power supplies vary rapidly in time, motivating the need for additional balancing resources~\cite{budischak2013cost}.

Since some peaking power plants may take more than an hour to bring online, during times of extensive peak usage, direct load control (i.e., load shedding) has been employed to ensure the security of the power system~\cite{molina2011decentralized}. 
However, the internet-connected distributed energy resources (DERs) are flexible in power demand and can be coordinated to provide ancillary services to the grid~\cite{meyn2015ancillary}.
Although the main idea underlying modern demand coordination has existed for decades~\cite{schweppe1980homeostatic}, the infrastructure required for load coordination is still in early stages, but developing rapidly~\cite{almassalkhi2018asynchronous,meyn2015ancillary,mathieu2013state}. 
Packetized energy managment (PEM) introduced previously by the authors~\cite{almassalkhi2018asynchronous,Kate2019}, is one such load coordination scheme. PEM leverages protocols used to manage data packets in communication networks to regulate the aggregate power consumption of DERs. More specifically, as in digital communication systems that break data into packets before transmission, PEM enables load control devices to consume energy in the form of "energy packets" which devices request periodically using a carefully designed randomized control policy. 
In PEM, the load coordinator only needs to know the aggregate power consumption and aggregate requests from the packetized-load to provide ancillary services to the grid. The energy-packet mechanism of PEM, therefore, provides a significant advantage in terms of communication overhead, over state-estimation based approaches, that require an entire histogram of states, which is addressed through observer design. Furthermore, controller complexity decreases in PEM, since the load coordinator only responds to individual requests depending upon the available flexibility as compared to more complex controllers.
By leveraging protocols that are similar to TCP/IP, PEM inherits certain properties with regard to providing statistically uniform access to the grid. PEM guarantees the quality of service (QoS) for individual DERs in the entire population through its unique opt-out mechanism. The mean-field approaches, on the other hand, ensure QoS in the mean sense of the population where individual DERs might violate the QoS~\cite{meyn2015ancillary}. 
This work describes how aggregated PEM resources can be coordinated in real-time and demonstrates the applicability of the method to practical power systems applications and the role of cyber-physical systems (CPS).


Historically, balancing authorities maintain real-time supply/demand balance through automatic generation control (AGC) and load-frequency control (LFC) by implementing PI controllers in steam turbine generator systems to ensure power system operation at nominal frequency~\cite{kundur1994power}. As the amount and distribution of controllable resources increases, determining an appropriate response to unscheduled events (e.g. power imbalances due to prediction error) is more challenging for the grid operators who need new tools for decision making.

An increasing number of researchers~\cite{kardakos2016optimal,mnatsakanyan2015novel} and industry groups~\cite{Sandia} are employing virtual power plants (VPPs) to aggregate groups of DERs and then dispatch those resources into energy markets, such as frequency regulation/AGC. VPPs are formed from aggregation of flexible resources which are limited in power and energy. Since PI-control-based AGC does not take energy state estimation of VPPs into account, it may overuse the offered flexibility in short period of time (greedy) which leads to the sudden saturation of VPPs (i.e. cannot provide any more flexibility). To overcome this phenomenon, model-predictive control (MPC) can be employed. MPC is a multi-input, multi-output (MIMO), optimization-based, predictive control technique that considers system constraints explicitly~\cite{borrelli2017predictive}. MPC strategies have previously been applied in power systems for optimal coordination of controllable loads, load shedding, capacity switching, tap-changer operation, etc. The main purpose of those strategies is contingency management, voltage stability, thermal control of transmission lines, and energy management~\cite{amini2018tradingoff,hiskens2005mpc,amini2018corrective}. In this work, an MPC scheme is employed to track a secure, economically-optimal reference trajectory of generators and VPPs while responding to power imbalances and satisfying physical constraints of the power system. For frequency regulation in the power system under high penetration of renewables, MPC has several advantages over PI controllers including robustness of the system against disturbance and uncertainty~\cite{ersdal2016model}.

This paper demonstrates the benefit that MPC has on dispatching resources with limited energy supply. An HIL platform is developed that consists of a transmission grid, MPC corrective dispatch scheme and PEM-enabled DERs emulated on a high-performance PC that requests packets of energy from the aggregator. The VPPs are physically realized in a micro-controller that connects the DERs to the grid via analog signals. The experimental results demonstrate the effectiveness of the MPC in a real-time CPS, thereby validating the ability of a VPP to track challenging signals under such control.


The remainder of the paper is organized as follows. Section II details Packetized Energy Management of DERs. In Section III, we demonstrate our cyber-physical validation platform. Section IV gives an overview of the system operational and control. Implementation results are provided in Section V and Section VI concludes the paper.

\section{Packetized energy management of DERs}
Packetized energy management is a bottom-up DER coordination scheme in which the DERs submit randomized requests of energy packets. The VPP accepts or rejects these requests based on the available flexibility. The DERs considered in this work are thermostatically controlled loads (TCLs) and energy storage systems (ESS).

The PEM-enabled DERs are designed to operate in one of the four following logical states: (i) charge (ii) discharge (iii) off (iv) opt-out. The first three states (charge, discharge, off) are associated with the normal PEM operation whereas the fourth OPT-OUT state ensures quality of service (QoS). A DER in OFF stochastically requests a charge packet or a discharge packet. If a charge packet is accepted, the DER changes state from OFF to CHARGE and consumes power for a specific time interval $\delta_c$. If a discharge packet is requested and consequently accepted, the DER transitions from OFF to DISCHARGE state and discharges power into the system for a fixed time $\delta_d$. After completing a charge or discharge packet, the device automatically transitions to OFF mode and this process of stochastically requesting charge/discharge packets repeats. PEM aims to maintain the DER's state within minimum and maximum operating bounds. PEM provides QoS guarantees by enabling the devices to opt out of the \emph{packetizing} behavior when the energy state goes outside of allowable upper and lower limits.

\section{Cyber-Physical Layout}

The smart grid paradigm \cite{NISTsmartgrid2014} is largely about the transformation of power systems into full cyber-physical systems that enable bidirectional flows of energy and communications. CPS are of vital importance to the grid, especially when increasing the presence of renewable generation and smart devices, improving control \cite{7017600}, and adding resiliency to the system \cite{NationalInstituteofStandardsandTechnologySmartGridInteroperabilityPanel2010NIST-IRSecurity}. The validation of CPS require accurate models of both cyber and physical sub-systems (e.g. HIL systems communicating with one another over realistic communication protocols). In order to validate the proposed demand-side CPS scheme, a real-time HIL platform is developed consisting of a transmission system and packetized load. The OP5600 real-time digital simulator from OPAL-RT is used to simulate the HIL cyber-physical system. The OP5600 has a multi-core processor along with digital and analog I/O with the capability of interfacing to a network of PCs in order to simulate large models in real-time\footnote{Herein, real-time refers to timescales on the order of tens of milliseconds.}. The RT-Lab software allows the communication between a host PC and the target (OP5600) simulator such that a real-time physical model can run on the simulator while the controller would run on the PC where an operator could make adjustments when necessary.

\begin{figure}[h!]
    \centering
    \includegraphics[width=0.9\columnwidth]{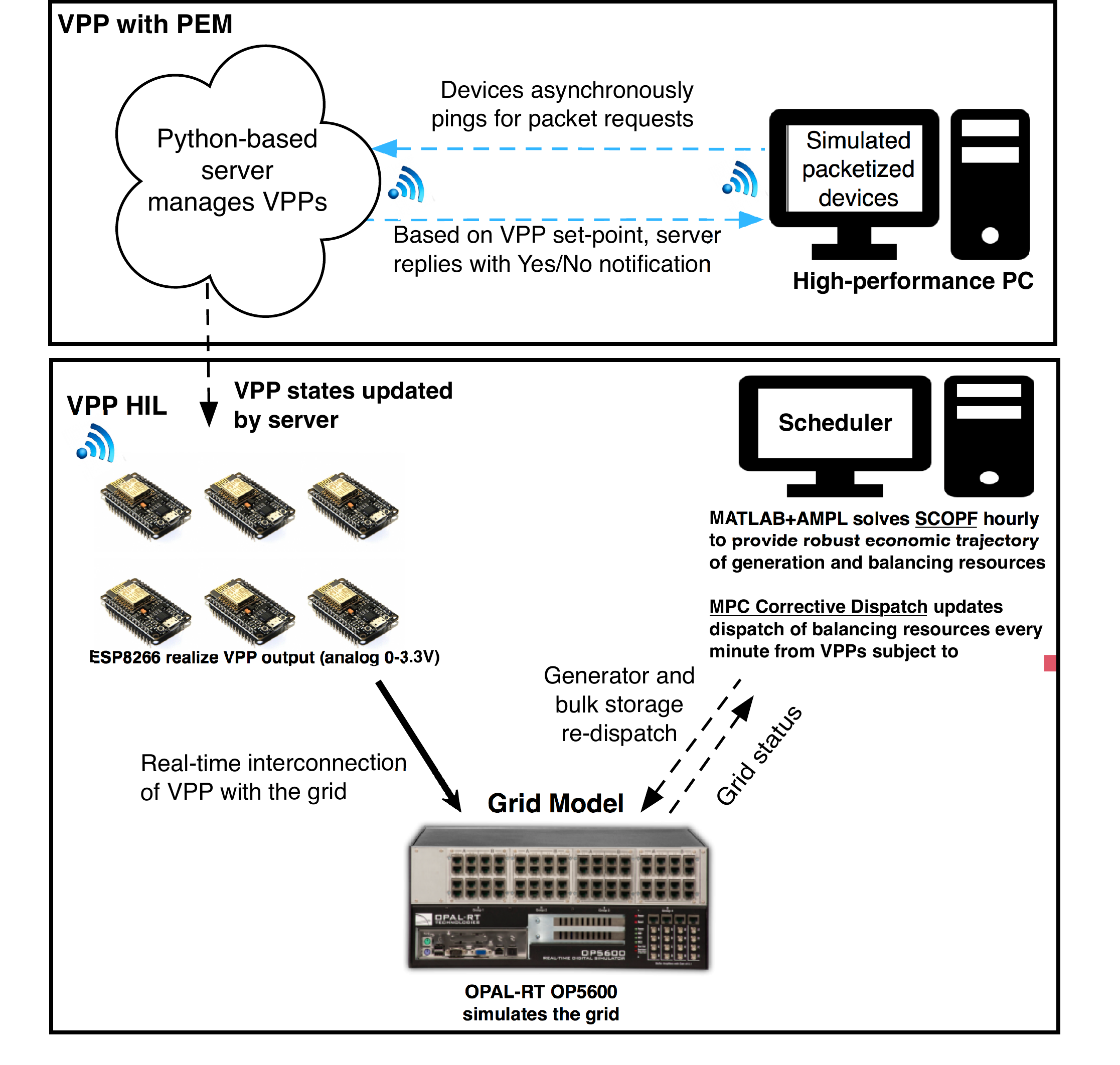}
    \caption{Cyber-physical platform overview: The transmission grid is simulated on the OP5600 and MPC-corrective dispatch is realized on a host PC and generates balancing signals. ESP8266 devices are connected to a python-based server via WiFi and transmit the VPPs' states through the analog interface. The packetized load is emulated on a high performance PC and requests energy packets from the VPP through WiFi communication.}
    \label{fig:fullsys_overview}
\end{figure}

``ePHASORSIM" is a tool developed by OPAL-RT to offer dynamic simulation of power systems in order to conduct power system studies and test control schemes. A grid is modeled with a standard positive-sequence equivalent single-phase constant-power AC model in ePHASORSIM, based on the Vermont Electric Power Company (VELCO) transmission system to be run in real time on the OP5600. RT-LAB and ePHASORSIM can be interfaced extremely easily with MathWorks' Simulink, which is used to develop the controls for the power system. The OPAL-RT blockset for Simulink allows a section of the Simulink block diagram to be run in real time on the OP5600 and the controls can be run asynchronously on the PC with the ability to accept user inputs when necessary. ESP8266 microchips were used to emulate VPP interconnections to the grid in real time. ESP8266 communicate with the cloud server over WIFI, while the server is being hosted on a Linux machine. Fig. \ref{fig:fullsys_overview} shows an overview of the cyber-physical platform used in this study.

\section{Energy aware dispatch of diverse energy \\resources}
\begin{figure}[t!]
      \centering
   \includegraphics[width=0.8\columnwidth]{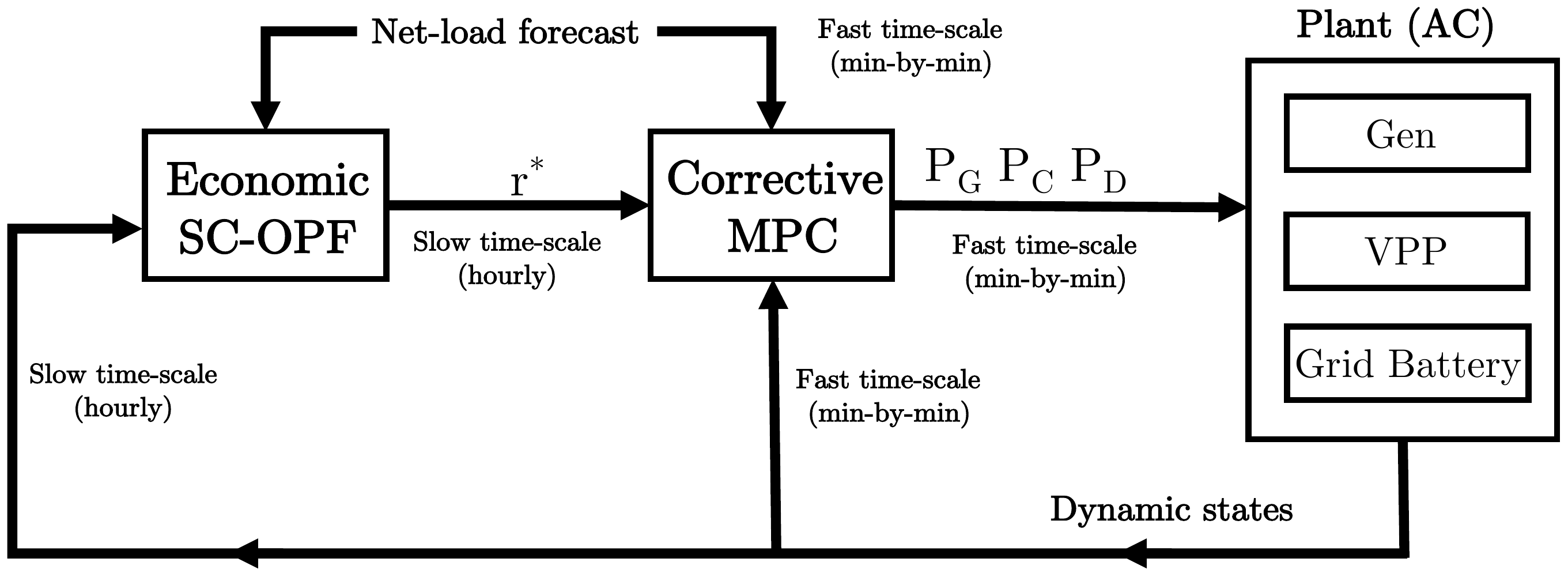}
      \caption{Overview of control scheme showing controller including OPF and MPC part and how each part is related to the power grid.}
      \label{fig:overview}
   \end{figure}

Security constrained optimal power flow (SCOPF) enables grid operators to implement economic schedules for generators, flexible loads and importing power into the area for a number of hours. However, the volatility and intermittent characteristic of net-loads (i.e., demand minus renewables) results in forecast error and power imbalances. Since grid operators may pay high penalties for rescheduling generators or importing power through tie lines to balance supply and demand~\cite{arnold2011model}, power mismatches can be balanced by controlling flexible resources. This suggests a bi-level control strategy where the first level is in charge of economic scheduling and its outputs are used as a reference input to the second level which is in charge of dispatching generators and VPPs to balance the system against any disturbance. An overview of the proposed control system is provided in Fig.~\ref{fig:overview}.


\subsection{AGC}

In the power system, safety of the electrical equipment and quality of delivered power is dependant on nominal system frequency. Therefore, the frequency should be controlled and monitored regularly and any mismatches in generation and consumption shall be corrected through load frequency control (LFC)~\cite{amini2016investigating}. Traditionally, the primary frequency regulation (speed-droop) on each generator stabilizes the power system with a steady-state frequency deviation from the desired system frequency depending on the droop characteristic and frequency sensitivity. A linear combination of frequency errors and change in imported power through tie lines from their scheduled contract basis is used as an error signal called area control error (ACE). AGC acts as a secondary control using an integral controller that sends  out control signals to generators and VPPs to reduce ACE to zero in steady state. For the purposes of this work, only two areas are used for simplicity, while being effective enough to demonstrate the flow of power between different areas. The first area represents a small balancing authority (control area), and the second represents the aggregate dynamics of the external system. One machine exists in the external area and has a large inertia and capacity to emulate the properties of the rest of the interconnected power system. 

\begin{figure}[ht]
    \centering
    \includegraphics[width=3.3in]{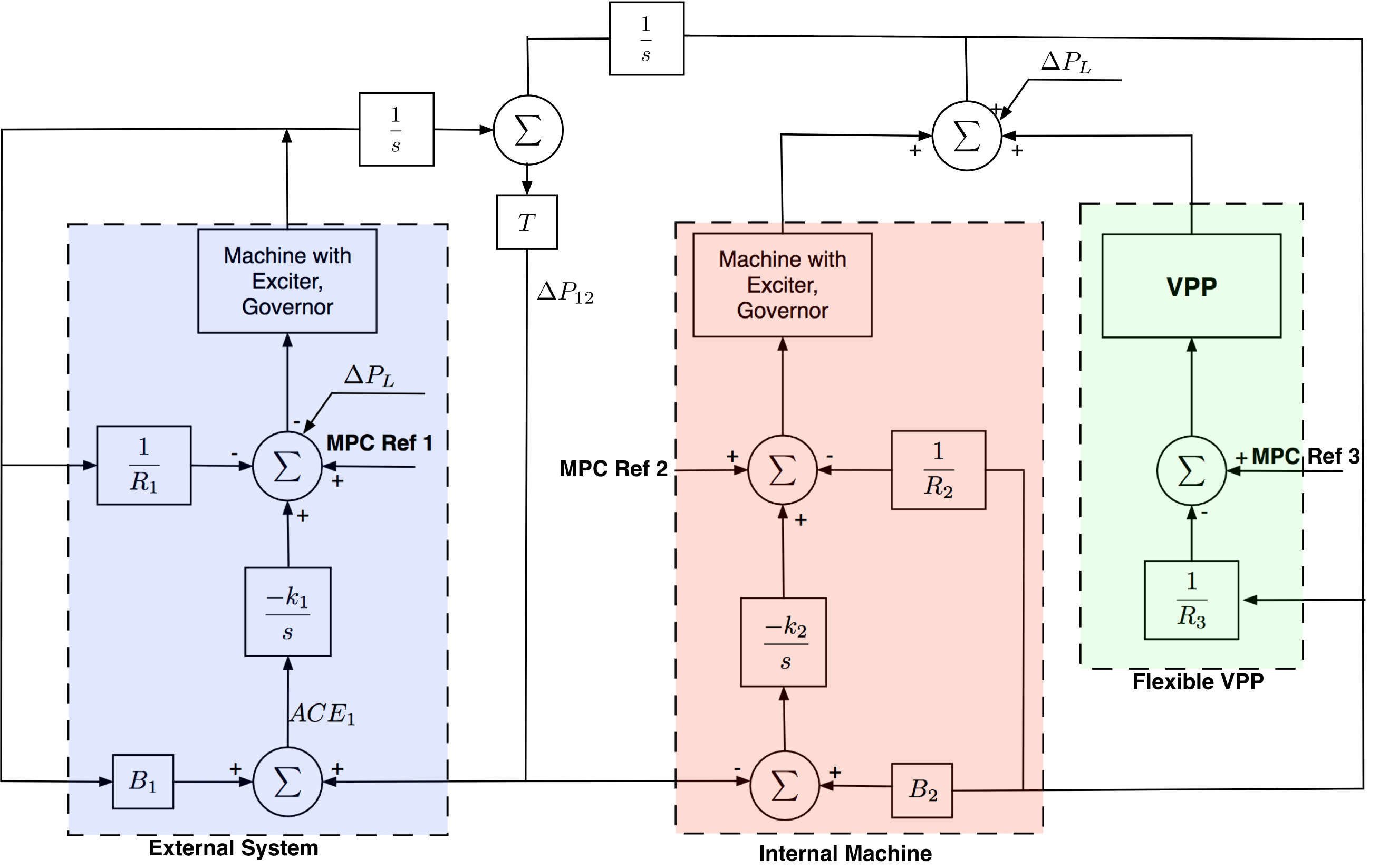}
    \label{fig:AGC_ControlDiagram}
    \caption{Diagram showing a control schematic for the test system including all of the generation in the internal and external areas.}
\end{figure}

Fig. 3 shows the control diagram for the system modeled in this paper, which is an adapted version of the diagram from \cite{kundur1994power}. The interaction between the internal and external areas involved are shown. The external machine, one of the internal machines, and the VPPs assist in AGC/ACE, while all of these generation sources are equipped for primary frequency regulation. 



\subsection{Model predictive control for power system}
Unlike conventional generators, VPPs (synthetic reserves) are energy-constrained and should be utilized considering their available flexibility. Aforementioned primary and secondary frequency controllers do not take energy states of the VPPs into account. Therefore, VPPs may reach their  energy capacity limits (saturate) unexpectedly and cannot provide balancing power anymore. As a result, conventional generators must be rescheduled to provide the required balancing services, which can be expensive, or even infeasible leading to a reliability risk.

As an advanced control technique, MPC forms an alternative to the PI controllers in frequency regulation which uses a mathematical model of the power system based on the current and future information and constraints to find the optimal control actions with respect to the defined objective. Unlike a PI controller, MPC dispatches resources (generators and VPPs) at each time step considering current states and forecasted conditions while handling the energy constraints of the flexible resources. The results of this work show that the MPC is more suitable for frequency regulation and dispatching energy-constrained resources compared to the PI controllers~\cite{hermans2012assessment}. The MPC scheme can be summarized as follows:
\begin{enumerate}
\item Controller uses measured/estimated initial states to solve an open-loop optimal control problem for $M$ steps, which is known as prediction horizon, taking into account current and future constraints. This gives a sequence of optimal open loop control actions and predict output. 
\item Apply receding horizon control so that only the first instance of the control sequence is given as the input to the plant.
\item Measure the actual system state after applying the first control action.
\item Go to step $1$.
\end{enumerate}

We consider a transmission system  model comprising of $N_b$ buses, $N_l$ lines,  $N_G$ generators, $N_L$ loads and $N_B$ VPPs. Parameters  $\Omega_i^N$ and $\Omega_i^G$ refer to a set of all buses connected to bus $i$ and set of all generators at bus $i$ respectively. Since MPC relies on a linear model of the actual system, the dynamic model of the system is discretized by forward Euler method with sample time $T_s$. The MPC optimization is defined to minimize the cost of the deviation of generator outputs from the scheduled set-points $P_{G,i}^r$ considering deviation cost $c_{G,i}$

\small
\begin{mini!}[3]
  {P_{G},P_{\text{ch}},P_{\text{dis}}}{ \sum_{l = k}^{k+M} \sum_{\forall i \in N_g} c_{G,i} ( P_{G,i}[l]-P_{G,i}^r[l] )^2  }
   {\label{eq:optVPP}}{J^*=}
   \addConstraint{\mkern-48mu P_{\text{ch},i}[l]-P_{\text{dis},i}[l]+P_{L,i}^f[l] + \sum_{j \in \Omega_i^N}f_{ij}[l] }{= \sum_{z \in \Omega_i^G}P_{G,z}[l] 
   \label{eq:powerbalance}}
   \addConstraint{\mkern-48mu f_{ij}[l] }{ = b_{ij}(\theta_i[l] - \theta_j[l]) \label{eq:lineflow}}
   \addConstraint{\mkern-48mu -\overline{f_{ij}} \leq f_{ij}[l]}{ \leq \overline{f_{ij}} \label{eq:linelimit} }
   \addConstraint{\mkern-48mu \underline{P_{G,i}}\leq P_{G,i}[l]}{ \leq \overline{P_{G,i}} \label{eq:Glim}} 
            \addConstraint{\mkern-48mu  -T_sR_{G,i} \leq P_{G,i}[l+1] - P_{G,i}[l]}{ \leq T_sR_{G,i} \label{eq:Gramp}}
       \addConstraint{\mkern-48mu 0 \leq P_{\text{ch},i}[l]}{ \leq \overline{P_{\text{ch},i}} \label{eq:Clim}} 
     \addConstraint{\mkern-48mu 0 \leq P_{\text{dis},i}[l]}{ \leq \overline{P_{\text{dis},i}} \label{eq:Dlim}} 
    \addConstraint{\mkern-48mu -T_sR_{\text{ch},i} \leq P_{\text{ch},i}[l+1] - P_{\text{ch},i}[l]}{ \leq T_sR_{\text{ch},i} \label{eq:Cramp}}
    \addConstraint{\mkern-48mu -T_sR_{\text{dis},i} \leq P_{\text{dis},i}[l+1] - P_{\text{dis},i}[l]}{ \leq T_sR_{\text{dis},i} \label{eq:Dramp}}
    \addConstraint{\mkern-48mu  S_i[l+1]}{ = S_i[l] + T_s\bigg(\eta_{\text{ch},i} P_{\text{ch},i}[l] - \eta_{\text{dis},i}^{-1} P_{\text{dis},i}[l]\bigg) \label{eq:dynSOC}}
     \addConstraint{\mkern-48mu \underline{S_i} \leq S_i[l]}{ \leq \overline{S_i} \label{eq:Slim}}
\end{mini!}
\normalsize
 where~\eqref{eq:powerbalance} imposes Kirchhoff's laws, implying that the net flow into a bus must equal the net flow out of that bus. Power flows on the line connecting bus $i$ and $j$ that are determined by~\eqref{eq:lineflow} must be within the power carrying capacity of the transmission line $\overline{f_{ij}}$ as shown in~\eqref{eq:linelimit}. Generators may inject power, $P_G$ and loads may consume power $P_L$ at each node $i$. Each conventional generator is described by its production state, which must be within generator upper and lower limits, $\overline{P_G}$ and $\underline{P_G}$, as shown in~\eqref{eq:Glim}. Furthermore, due to the thermal nature of the generators, their ramp rates are limited to up and down limits, $R_G$,  as shown in~\eqref{eq:Gramp}. The responsive VPPs overcome limitations of generator ramping rates. Non-negative scalar  $P_{\text{ch}}$ and $P_{\text{dis}}$ represent charging and discharging power of a VPP. The charging and discharging efficiencies are denoted  by $\eta_{\text{ch}}$ and $\eta_{\text{dis}}$. Charging and discharging power and SOC of the VPPs are subject to constraints~\eqref{eq:Clim},~\eqref{eq:Dlim} and~\eqref{eq:Slim} where $\overline{P_{\text{ch}}}$, $\overline{P_{\text{dis}}}$, $\overline{S}$, and $\underline{S}$ represent maximum charging and discharging power and the maximum and minimum energy capacity of VPP, respectively. In general, coordination schemes do not offer instant control over all DERs in a fleet, but are subject to separate internal control, actuation, and communication loops~\cite{Duffaut2018tpwrsP2}. These cyber-physical control considerations manifest themselves as 
ramp-rate limits on the charging ($R_{\text{ch}}$) and discharging ($R_{\text{dis}}$) of VPPs as shown in~\eqref{eq:Cramp} and~\eqref{eq:Dramp}. The dynamic of the VPP's SOC is shown in~\eqref{eq:dynSOC}. 

\section{Results}
The section experimentally demonstrates that energy-aware dispatch of flexible VPPs enhances the AGC performance. The details of the test-setup are as follows. The transmission system consisting of 161 buses, 223 transmission lines and three generating units, is set up in ePHASORSIM and supplies a total load of 609 MW consisting of approximately $50\%$ renewable generation and remaining load is supplied from one external and two internal machines and generators. The flexibility is provided by two VPPs consisting of one bulk battery and one HIL VPP. The HIL VPP consists of real packetized-enabled DERs emulated on a high-performance PC, that requests the VPP (server) for packets of energy through web-sockets. The VPP obtains balancing signals from the grid operator and accepts/rejects the packets based on the available flexibility. ESP8266 is the physical realization of VPP that obtains the VPP's state from the server and sends it to the grid through an analog interface.

\subsection{Capacity saturation of VPP}
\begin{figure}[ht!]
      \centering
   \includegraphics[width=0.9\columnwidth]{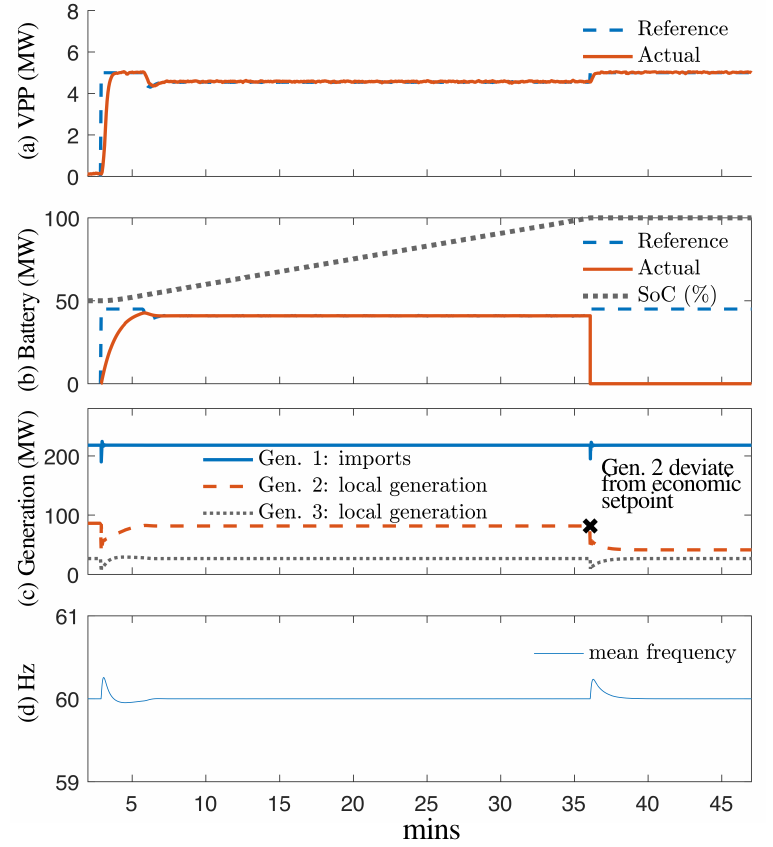}
      \caption{(a) The HIL VPP's actual and reference power (MW) (b) Grid scale battery's actual power (MW), reference power (MW) and state of charge (SOC $\%$) during charge/discharge events (c) Generators' power output (MW) (d) Generators' mean frequency (Hz). The saturation of the VPP to a step decrease in load is shown in this figure. For the change in load, the HIL VPP and the battery charges at a continuous rate. The battery saturates at about t = $36$ mins, after which their output goes to zero and cannot support the requested flexibility.}
    \label{fig:HiL_AGC_sat}
\end{figure}

The HIL VPP and grid scale battery providing grid services are energy-limited and ignoring their energy capacity results in inferior AGC performance. The effect of their capacity on the ancillary services provided to the grid is demonstrated in Fig.~\ref{fig:HiL_AGC_sat}. The capacity of the battery is $45$ MWh. Fig.~\ref{fig:HiL_AGC_sat} shows that the load reduces by $50$ MW resulting in excess generation and both battery and VPP take up $45$ MW and $5$ MW respectively, of this excess generation. The battery is initially $50\%$ charged, as shown in Fig.~\ref{fig:HiL_AGC_sat} (b). After $t \approx 36$ mins, in Fig.~\ref{fig:HiL_AGC_sat} (b), the battery saturates and can no longer provide ancillary services to the grid. The power output of Gen.~2 is therefore reduced to account for the loss of $45$ MW and the system deviates from its scheduled generation (Fig.~\ref{fig:HiL_AGC_sat} (c)) in order to keep the system stable (Fig.~\ref{fig:HiL_AGC_sat} (d)). 

\subsection{MPC with capacity saturation}
\begin{figure}[ht!]
      \centering
   \includegraphics[width=0.9\columnwidth]{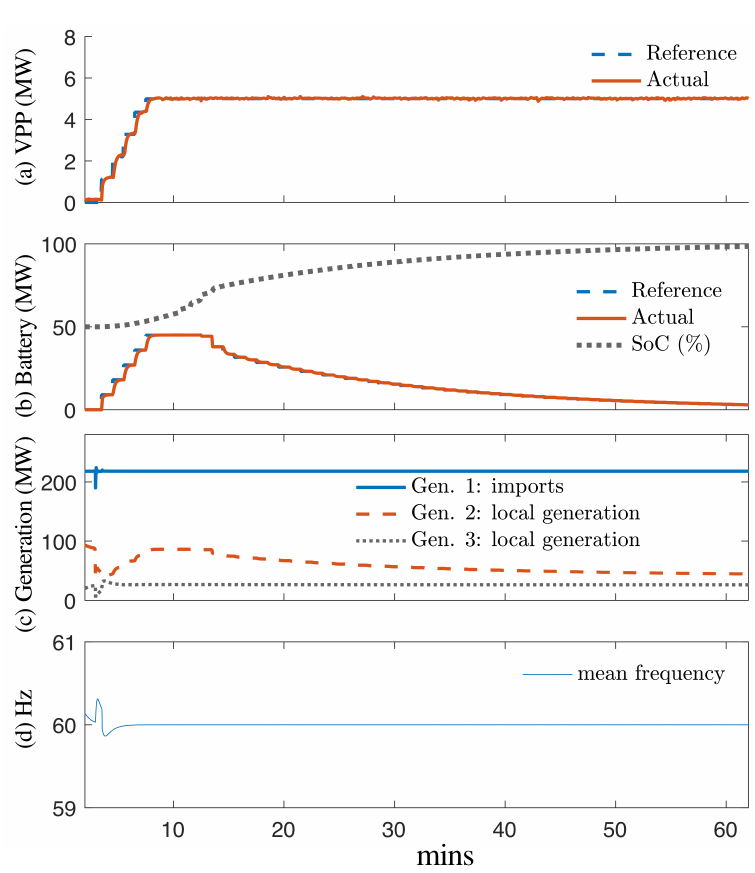}
    \caption{(a) The HIL VPP's actual power (MW) and reference power (MW) (b) Grid scale battery's actual power (MW), reference power (MW) and state of charge (SOC $\%$) during charge/discharge events (c) Generators' power output (MW) (d) Generators' mean frequency (Hz). MPC with capacity saturation takes into consideration the current state of charge of the VPP and initially ramps up to the requested $50$ MW. However, at t = $13$ mins, MPC lowers the setpoint in steps to avoid VPP saturation and provide support to the system for a longer time. }
    \label{fig:HiL_MPC}
\end{figure}

The proposed \emph{energy aware} scheme improves the performance by explicitly accounting for the energy limits of the flexible resources. Fig.~\ref{fig:HiL_MPC} shows a load change $\Delta P_L = 50$ MW that results in the HIL VPP and the battery providing the remaining slack. MPC keeps track of the current state of charge of the HIL VPP and the battery (Fig.~\ref{fig:HiL_MPC} (a) and (b)) and after $t \approx 13$ mins, gradually reduces the set-points of the battery to avoid saturation as shown, in Fig.~\ref{fig:HiL_MPC} (b). The HIL VPP and the battery can therefore supply ancillary services to the grid for over $60$ mins.

\section{Discussion/Conclusion}
This paper presents a hardware-in-the-loop implementation of PEM-based cyber-physical platform and demonstrates that aggregated PEM-enabled DERs can 
provide ancillary services to the grid. The system consists of emulated DERs, an aggregator realized as real live webserver, and a transmission system developed from the real data provided by VELCO. The experimental studies carried out in this work show that conventional control schemes (i.e. AGC and droop) do not take into account the state of charge of VPPs while providing ancillary services to the grid that leads to capacity saturation of the VPPs. As a result, the VPP can no longer support the requested flexibility and cause a disturbance in the grid because the generators have to ramp-up quickly to ensure stability. Conventional schemes are therefore greedy in managing flexible resources and leads to capacity saturation and unwanted disturbances. Model predictive control (MPC) responds to the unexpected disturbances by dispatching resources depending upon the hourly load forecast as well as minute-by-minute dispatch in a receding horizon. MPC is shown to proactively vary the consumption of flexible resources depending upon the SOC, which utilizes these resources for a longer time than the base case with AGC. The generators therefore operate as close as possible to their optimal limits for a much longer time. 
\vspace{-0.1cm}
\section*{Acknowledgment}
The authors would like to extend thanks to Bernadette Ferndandez and Chris Root of VELCO for providing actual transmission and bus data for the Vermont system to help provide a more realistic study.

\vspace{-0.1cm}
{\small
\bibliographystyle{IEEEtran}
\bibliography{ref} 

\begin{thebibliography}{10}
\providecommand{\url}[1]{#1}
\csname url@samestyle\endcsname
\providecommand{\newblock}{\relax}
\providecommand{\bibinfo}[2]{#2}
\providecommand{\BIBentrySTDinterwordspacing}{\spaceskip=0pt\relax}
\providecommand{\BIBentryALTinterwordstretchfactor}{4}
\providecommand{\BIBentryALTinterwordspacing}{\spaceskip=\fontdimen2\font plus
\BIBentryALTinterwordstretchfactor\fontdimen3\font minus
  \fontdimen4\font\relax}
\providecommand{\BIBforeignlanguage}[2]{{%
\expandafter\ifx\csname l@#1\endcsname\relax
\typeout{** WARNING: IEEEtran.bst: No hyphenation pattern has been}%
\typeout{** loaded for the language `#1'. Using the pattern for}%
\typeout{** the default language instead.}%
\else
\language=\csname l@#1\endcsname
\fi
#2}}
\providecommand{\BIBdecl}{\relax}
\BIBdecl

\bibitem{budischak2013cost}
C.~Budischak \emph{et~al.}, ``Cost-minimized combinations of wind power, solar
  power and electrochemical storage, powering the grid up to 99.9\% of the
  time,'' \emph{Journal of Power Sources}, vol. 225, pp. 60--74, 2013.

\bibitem{molina2011decentralized}
A.~Molina-Garcia \emph{et~al.}, ``Decentralized demand-side contribution to
  primary frequency control,'' \emph{IEEE Transactions on Power Systems},
  vol.~26, no.~1, pp. 411--419, Feb 2011.

\bibitem{meyn2015ancillary}
S.~P. Meyn \emph{et~al.}, ``Ancillary service to the grid using intelligent
  deferrable loads,'' \emph{IEEE Transactions on Automatic Control}, vol.~60,
  no.~11, pp. 2847--2862, 2015.

\bibitem{schweppe1980homeostatic}
F.~C. Schweppe \emph{et~al.}, ``Homeostatic utility control,'' \emph{IEEE
  Transactions on Power Apparatus and Systems}, no.~3, pp. 1151--1163, 1980.

\bibitem{almassalkhi2018asynchronous}
M.~Almassalkhi \emph{et~al.}, ``Asynchronous coordination of distributed energy
  resources with packetized energy management,'' \emph{\textnormal{in} Energy
  Markets and Responsive Grids}, pp. 333--361, Springer, 2018.

\bibitem{mathieu2013state}
J.~L. Mathieu \emph{et~al.}, ``State estimation and control of electric loads
  to manage real-time energy imbalance,'' \emph{IEEE Transactions on Power
  Systems}, vol.~28, no.~1, pp. 430--440, 2013.

\bibitem{Kate2019}
K.~Desrochers \emph{et~al.}, ``Real-world, full-scale validation of power
  balancing services from packetized virtual batteries,'' \emph{IEEE PES
  Innovative Smart Grid Technologies Conference (ISGT)}, 2019.

\bibitem{kundur1994power}
P.~Kundur \emph{et~al.}, \emph{Power system stability and control}.\hskip 1em
  plus 0.5em minus 0.4em\relax McGraw-hill New York, 1994, vol.~7.

\bibitem{kardakos2016optimal}
E.~G. Kardakos \emph{et~al.}, ``Optimal offering strategy of a virtual power
  plant: A stochastic bi-level approach,'' \emph{IEEE Transactions on Smart
  Grid}, vol.~7, no.~2, pp. 794--806, 2016.

\bibitem{mnatsakanyan2015novel}
A.~Mnatsakanyan and S.~W. Kennedy, ``A novel demand response model with an
  application for a virtual power plant,'' \emph{IEEE Transactions on Smart
  Grid}, vol.~6, no.~1, pp. 230--237, 2015.

\bibitem{Sandia}
J.~Johnson \emph{et~al.}, ``Design and evaluation of a secure virtual power
  plant,'' \emph{Sandia Technical Report, SAND2017-10177}, 09 2017.

\bibitem{borrelli2017predictive}
F.~Borrelli \emph{et~al.}, \emph{Predictive control for linear and hybrid
  systems}.\hskip 1em plus 0.5em minus 0.4em\relax Cambridge University Press,
  2017.

\bibitem{amini2018tradingoff}
M.~Amini and M.~Almassalkhi, ``Trading off robustness and performance in
  receding horizon control with uncertain energy resources,''
  \emph{\textnormal{in} IEEE Power Systems Computation Conference}, pp. 1--7,
  June 2018.

\bibitem{hiskens2005mpc}
I.~A. Hiskens and B.~Gong, ``Mpc-based load shedding for voltage stability
  enhancement,'' \emph{Proceedings of the 44th IEEE Conference on Decision and
  Control}, pp. 4463--4468, Dec 2005.

\bibitem{amini2018corrective}
M.~Amini and M.~Almassalkhi, ``Corrective dispatch of uncertain energy
  resources using chance-constrained receding horizon control,'' \emph{arXiv
  preprint arXiv:1810.08685}, 2018.

\bibitem{ersdal2016model}
A.~M. Ersdal \emph{et~al.}, ``Model predictive load-frequency control,''
  \emph{IEEE Transactions on Power Systems}, vol.~31, no.~1, pp. 777--785,
  2016.

\bibitem{NISTsmartgrid2014}
\BIBentryALTinterwordspacing
NIST, \emph{NIST Framework and Roadmap for Smart Grid Interoperability
  Standards, Release 3.0}, Sep 2014. [Online]. Available:
  \url{https://www.nist.gov/sites/default/files/documents/smartgrid/NIST-SP-1108r3.pdf}
\BIBentrySTDinterwordspacing

\bibitem{7017600}
S.~Xin \emph{et~al.}, ``Cyber-physical modeling and cyber-contingency
  assessment of hierarchical control systems,'' \emph{IEEE Transactions on
  Smart Grid}, vol.~6, no.~5, pp. 2375--2385, Sept 2015.

\bibitem{NationalInstituteofStandardsandTechnologySmartGridInteroperabilityPanel2010NIST-IRSecurity}
\BIBentryALTinterwordspacing
{National Institute of Standards and Technology Smart Grid Interoperability
  Panel}, \emph{{NIST-IR 7628 Volume 1, Guidelines for Smart Grid Cyber
  Security}}, 2010, no. September. [Online]. Available:
  \url{http://permanent.access.gpo.gov/gpo1900/nistir-7628_total.pdf}
\BIBentrySTDinterwordspacing

\bibitem{arnold2011model}
M.~Arnold and G.~Andersson, ``Model predictive control of energy storage
  including uncertain forecasts,'' \emph{\textnormal{in} Power Systems
  Computation Conference, Stockholm, Sweden}, vol.~23, pp. 24--29, Aug 2011.

\bibitem{amini2016investigating}
M.~Amini and M.~Almassalkhi, ``Investigating delays in frequency-dependent load
  control,'' \emph{\textnormal{in} IEEE Innovative Smart Grid Technologies -
  Asia}, pp. 448--453, Nov 2016.

\bibitem{hermans2012assessment}
R.~M. Hermans \emph{et~al.}, ``Assessment of non-centralised model predictive
  control techniques for electrical power networks,'' \emph{International
  journal of control}, vol.~85, no.~8, pp. 1162--1177, 2012.

\bibitem{Duffaut2018tpwrsP2}
L.~Duffaut~Espinosa \emph{et~al.}, ``{A Packetized Energy Management
  Macro-model, Part II: Tracking with Diverse DER Populations},'' Submitted to
  the IEEE Transactions on Power Systems.

\end{thebibliography}
}
%










\end{document}